\documentclass[12pt,preprint]{aastex}

\newcommand{\etal}{{\em et al.}}            

\shorttitle{Cen A lobe}
\shortauthors{Kraft \etal}

\begin{document}

\title{A {\em Chandra} Study of the Lobe/ISM Interactions Around the Inner Radio Lobes of Centaurus A:  Constraints on the Temperature Structure and Transport Processes}
\author{R. P. Kraft, P. E. J. Nulsen}
\affil{Harvard/Smithsonian Center for Astrophysics, 60 Garden St., MS-67, Cambridge, MA 02138}
\author{M. Birkinshaw, D. M. Worrall}
\affil{University of Bristol, Department of Physics, Tyndall Avenue, Bristol BS8 ITL, UK}
\author{R. F. Penna}
\affil{Harvard/Smithsonian Center for Astrophysics, 60 Garden St., Cambridge, MA 02138 and Department of Physics and Astronomy, University of Rochester, 500 Wilson Boulevard, 
Rochester, NY 14627}
\author{W. R. Forman}
\affil{Harvard/Smithsonian Center for Astrophysics, 60 Garden St., Cambridge, MA 02138}
\author{M. J. Hardcastle}
\affil{University of Hertfordshire, School of Physics, Astronomy, and Mathematics, Hatfield AL10 9AB, UK}
\author{C. Jones}
\affil{Harvard/Smithsonian Center for Astrophysics, 60 Garden St., Cambridge, MA 02138}
\author{S. S. Murray}
\affil{Harvard/Smithsonian Center for Astrophysics, 60 Garden St., Cambridge, MA 02138}

\begin{abstract}

We present results from deeper {\em Chandra} observations of the southwest
radio lobe of Centaurus A, first described by \citet{kra03}.
We find that the sharp X-ray surface brightness
discontinuity extends around $\sim$75\% of the periphery of the radio lobe, and
detect significant temperature jumps in the brightest regions of
this discontinuity nearest to the nucleus.  This demonstrates that this
discontinuity is indeed a strong shock which is the result of
an overpressure which has built up in the entire lobe over time.
Additionally, we demonstrate that if the mean free path for ions to transfer energy
and momentum to the electrons behind the shock
is as large as the Spitzer value, the electron and proton temperatures
will not have equilibrated along the SW boundary of the radio lobe
where the shock is strongest.
Thus the proton temperature of the shocked gas could be considerably larger
than the observed electron temperature, and the total energy of the outburst
correspondingly larger as well.
We investigate this using a simple one-dimensional shock model
for a two-fluid (proton/electron) plasma.
We find that for the thermodynamic parameters of the Cen A shock
the electron temperature rises rapidly from $\sim$0.29 keV (the temperature
of the ambient ISM) to $\sim$3.5 keV
at which point heating from the protons is balanced by adiabatic losses.
The proton and electron temperatures do not equilibrate in a timescale
less than the age of the lobe.
We note that the measured electron temperature of similar features
in other nearby powerful radio galaxies in poor environments
may considerably underestimate the strength and velocity of the shock.

\end{abstract}


\section{Introduction}

Radio galaxies are believed to evolve through three phases.  Initially,
the lobes surrounding the jets are greatly overpressured relative to the ambient medium
and the inflation of lobes is highly supersonic.  
The early, highly supersonic phase of lobe inflation is short-lived in
most sources and has been conclusively identified in only a small
number of radio galaxies and clusters of galaxies
including Centaurus A \citep{kra03} and NGC 3801 \citep{cro06}.
As the inflation continues and the bubbles become larger, 
the pressure in the lobes drops
and approaches equilibrium with the ambient gas.
In these systems, such as Hydra A \citep{nul05} and M87
\citep{for05}, the weak shock surrounding the lobe
is often observable as a surface brightness
discontinuity in the X-ray emission.  
Ultimately, the bubble loses energy (via adiabatic expansion and
perhaps thermal conduction) as it rises buoyantly in the atmosphere and becomes
effectively unobservable, although these late-stage bubbles (radio
relics) may become re-energized by mergers \citep{rey02,enss02}.

The proximity of Centaurus A \citep[d$\sim$3.4 Mpc, five times
closer than the Virgo cluster, see][]{isr98} makes it an ideal astrophysical laboratory.
Features can be observed with a sensitivity and linear resolution
unattainable in any other active galaxy, allowing detailed study
of the hydrodynamics and energetics of lobe inflation.  In our previous paper on the
X-ray emission from the southwest radio lobe of Centaurus A, we reported
the discovery of a hot ($\sim$3.5 keV) shell of X-ray emission surrounding
the lobe.  We interpreted this shell as the result of the highly
supersonic expansion/inflation ($M\sim$8) of the lobe into the ambient ISM.
The dynamics of this process are of great interest because they can yield information
on the transport physics of the ICM of clusters of galaxies and early-type galaxies,
and on the roles that viscosity and thermal conduction play
in the release of energy into cool cluster cores.

In this paper, we present results from an analysis of four {\em Chandra}
pointed observations of Centaurus A, focusing on the morphology
and temperature structure of the X-ray shell around
the SW radio lobe.  The combined observation time of the data presented
in this paper is 150 ks, more than double that used in the analysis of
\citet{kra03}, and the detector roll angle and pointing of the later
observations are better suited to study the lobe.
As a result we can study the details of the transport processes in the lobe
shock on scales previously observable only in Galactic or Magellanic supernova remnants.
We report two important new results.  First, the surface brightness discontinuity
between the SW radio lobe and the ISM extends around most of the periphery of
the lobe, not just the SW corner as reported by \citet{kra03}.  
We find marginal evidence for a temperature gradient
in the shocked gas across the X-ray bright enhancement at the southwestern
boundary of the radio lobe.  Second, we demonstrate that
if the thermal equilibration time of the electrons and
ions in the gas is as slow as the Spitzer rate, the electrons will not have
thermalized.  This suggests that the electron temperature inferred from the X-ray
spectra considerably underestimates the strength of the shock, as has been 
reported for several Galactic and Magellanic SNRs.
In addition, we detect sharp surface brightness discontinuities around the NE radio lobe,
but lack sufficient source counts to accurately determine its gas density an temperature.
The features have temperatures above 1 keV, and thus with their morphologies and
locations, are suggestive of shocks.

This paper is organized as follows.  Section two
contains a summary of the observational details.
We present the results of the data analysis in section 3, and
we discuss the implications in section 4.
Section 5 contains a brief summary and conclusions, as well
as possible future observations.
We assume a distance of 3.4 Mpc to Cen A \citep{isr98} for
consistency with our previous work.
At this distance, 1$''$=17 pc.  All uncertainties are at 90\%
confidence unless otherwise stated, and all coordinates are J2000.
All elemental abundances in this paper are relative to the Solar
abundances tabulated by \citet{angr}.

\section{Data Analysis and Methods}

Centaurus A has been observed four times with {\em Chandra}/ACIS, twice with
ACIS-I for $\sim$35 ks each in AO-1, and twice with
ACIS-S for $\sim$50 ks each in AO-3 and AO-4 at the same roll angle.
Results on the southwest lobe from the first two ACIS-I observations 
have been published in \citet{kra03}.  
The additional observations more than double the effective exposure.
Results from the additional observations on the jet have already been published \citep{mjh03,kat06}.
We filtered all data for periods of high background, and removed events
occurring at node boundaries.
The total good times of the ACIS-S and ACIS-I observations are $\sim$94 ks
and $\sim$68 ks, respectively.
The four data sets were coaligned relative to each other to better than
0.1$''$ by centroiding the positions of 30 bright X-ray binaries within 5$'$ of
the nucleus.  The absolute position was then fixed by aligning the radio
and X-ray centroid of the nucleus.
A comparison of the positions of X-ray binaries
and globular clusters demonstrates that the absolute sky coordinates are
accurate to better than 0.5$''$ (Woodley \etal~2007, submitted).
All four data sets are used for spectral analysis, but only the
two ACIS-S observations are used for images and surface brightness profiles presented
in this paper.  The advantage in signal to noise that might
be gained from combining the ACIS-S and ACIS-I observations 
is more than offset by the complexities
of interpreting the imaging analysis of data taken at different instrument rolls.
Cen A lies at relatively low Galactic latitude ($b$=19$^\circ$.4) and behind
the North Polar Spur.  The ACIS blank sky backgrounds, created from
multiple observations at high galactic latitude, are inappropriate for
these observations.  Local background is used for all spectral analysis.  

\section{Results}

\begin{figure}
\plotone{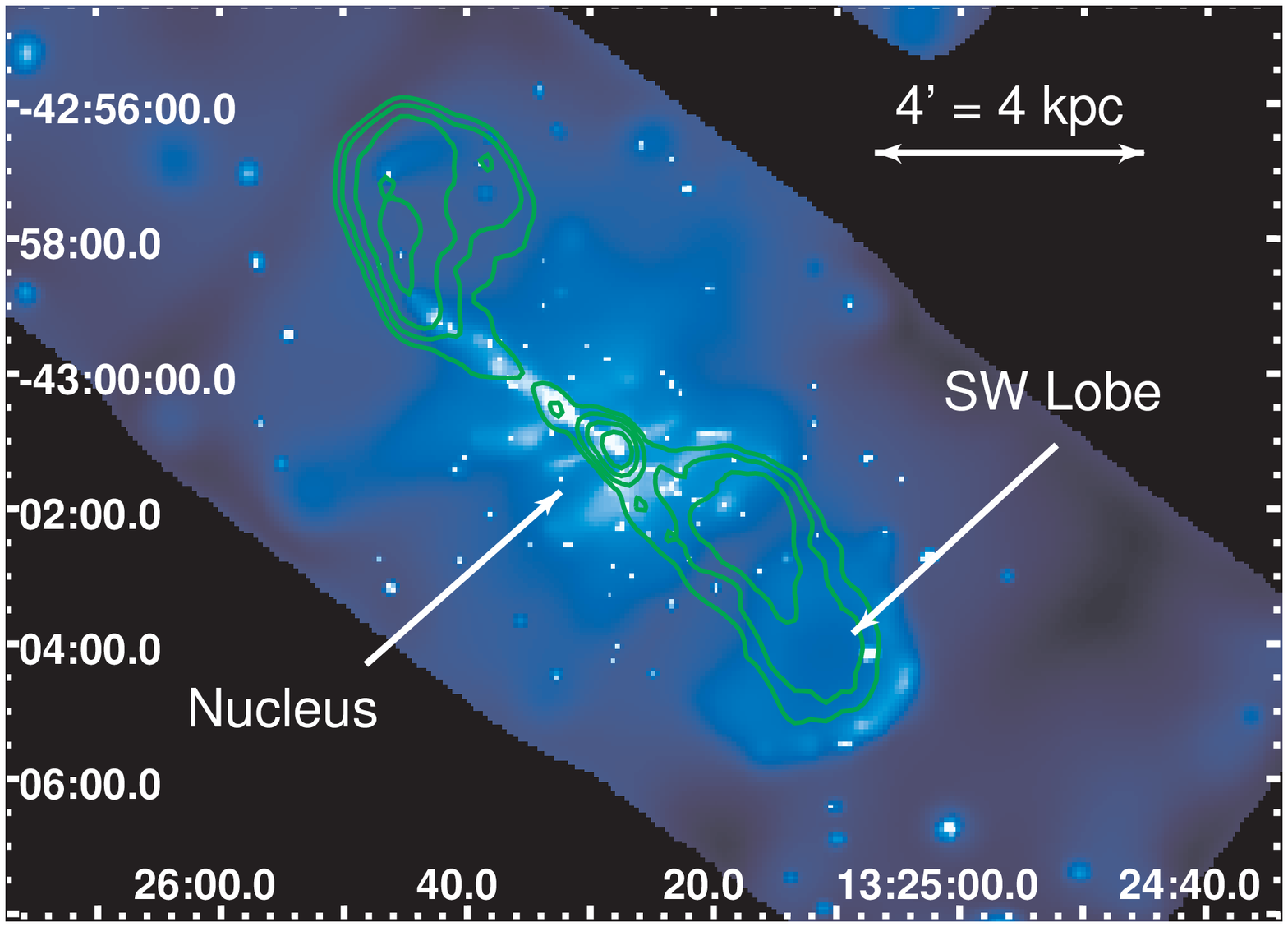}
\caption{Adaptively smoothed, exposure corrected, background subtracted Chandra/ACIS-S
image of Centaurus A in the 0.5-2.0 keV band.
Radio contours (13 cm - 30$''\times$20$''$ beam FWHM) are overlaid.}\label{xradovl}
\end{figure}

\begin{figure}
\plotone{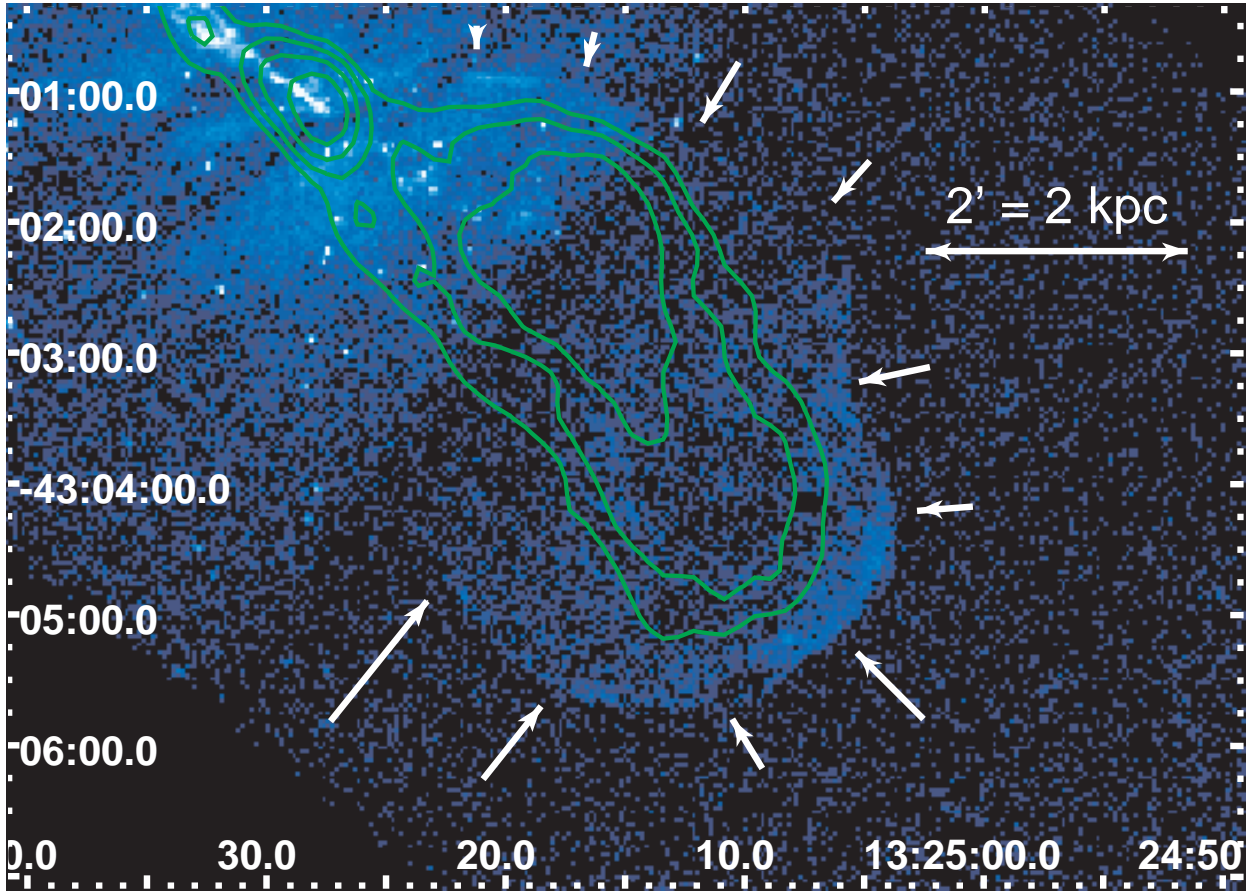}
\caption{Raw X-ray image of the southwest radio lobe of Centaurus A in
the 0.5-2.0 keV band.  Radio contours (13 cm - 30$''\times$20$''$ beam) are overlaid.
The white arrows denote the surface brightness discontinuity which
delineates the outer edge of the shock-heated shell of gas.}\label{xraw}
\end{figure}

An adaptively-smoothed, exposure-corrected, background subtracted 
X-ray image created from the two Chandra/ACIS-S observations in the 0.5-2.0 keV band,
with 13 cm radio contours overlaid, is shown in Figure~\ref{xradovl}.
It was not possible to remove all the detector artifacts from this image,
and the dark bands running NW/SE just beyond the NE lobe and through the middle
of the SW lobe are chip gaps.
A raw X-ray image in the same energy band is shown in Figure~\ref{xraw}.
An X-ray enhancement surrounds most of the lobe as
denoted by the white arrows in Figure~\ref{xraw}, and
is visible in both images.  In our previous paper, we found that the temperature
of this hot shell at the periphere of the SW lobe
is $\sim$3.5 keV.  Since the temperature of the ISM
is $\sim$0.3 keV \citep{kra03}, the inflation of the lobe is driving a strong shock into the ISM, at
least toward the SW.

\subsection{SW Radio Lobe}

The new, deeper Chandra observations of Centaurus A show details of
the structure of this high Mach number shock that
were not visible in the shorter ACIS-I observations.
First, it is clear from Figure~\ref{xraw} that the surface brightness
discontinuity between the ISM and shocked gas is visible around
$\sim$3/4 of the periphery of the lobe.  This suggests that the lobe
is inflating more or less spherically (i.e. energy dominated),
and is not simply being driven by jet ram pressure radially away 
from the nucleus (i.e. momentum dominated).
This is consistent with the fact that the minimum pressure of the radio
lobe greatly exceeds the pressure of the ISM \citep{kra03}.
The shock is strongest (in the sense that the electron temperature of
the post-shock gas is highest, $\sim$3.5 keV) at the southwestern edge of the lobe,
where the ambient gas density is lowest.

\begin{figure}
\plotone{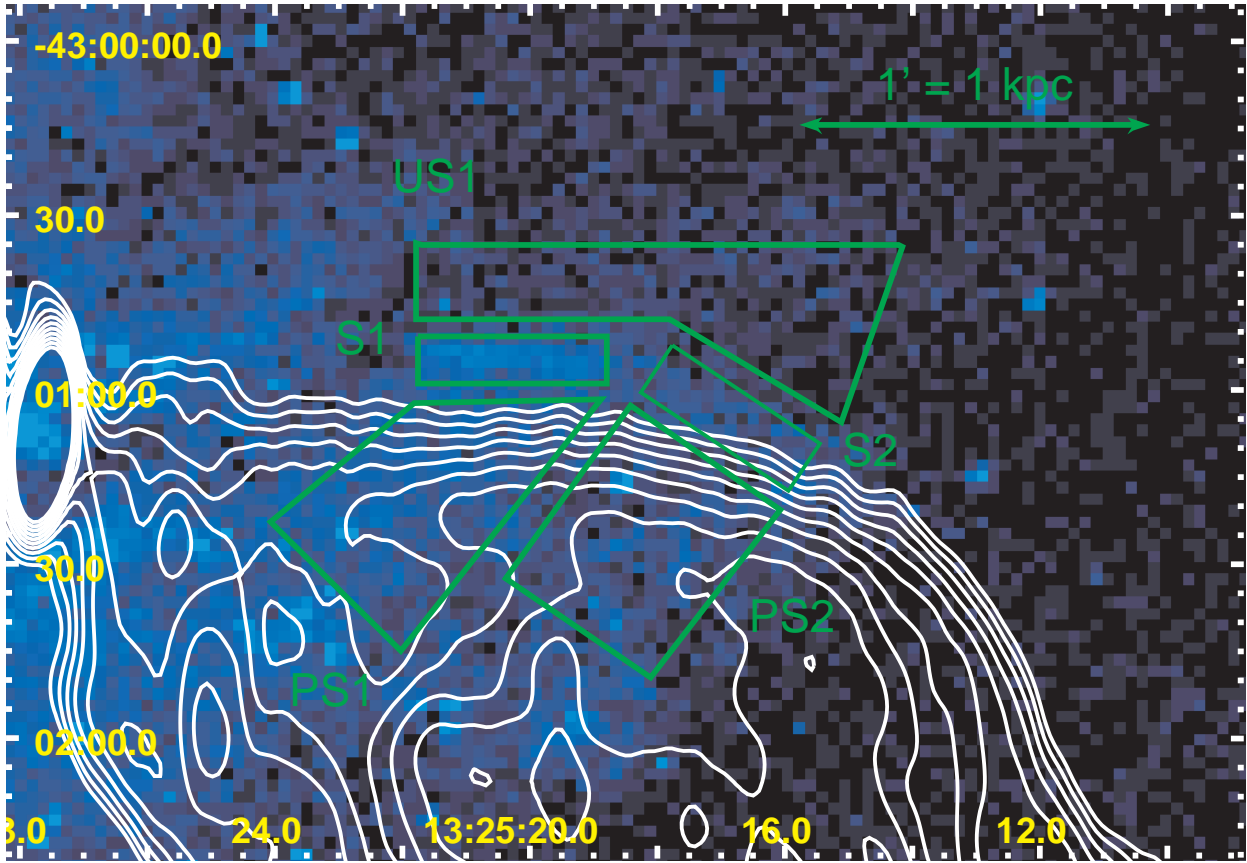}
\caption{Raw X-ray image of the southwest radio lobe of Centaurus A in
the 0.5-2.0 keV band showing the shock (regions S1 and S2) along the northern
periphery of the lobe.  Radio contours (1.54 GHz - 16.25$''\times$4.80$''$ beam) 
are overlaid.}\label{xrawz}
\end{figure}

The effect of the shock propagating in a region of denser gas can
clearly be seen in the vicinity of the northern periphery of
the lobe in Figures~\ref{xraw} and~\ref{xrawz}.
Sharp discontinuities in the X-ray surface brightness are labeled
S1 and S2.  The sharpness of these features strongly suggests that they are due 
to shocks being driven into the ISM by the lobe expansion.
S1 is roughly twice as bright as S2, but the
ISM behind S1 is also brighter than that behind S2.
Thus the X-ray surface brightness of the shocked gas is highest
where the ambient density of the ISM is highest.
In addition, [O III] emission lines have been detected
in this region (Joss Bland-Hawthorn, private communication, 2006),
suggesting that the lobe is shock-heating the multi-phase
ISM of the merging spiral galaxy.  The details of this will
be presented in a future publication.

We extracted spectra from five regions:  two rectangular regions
corresponding to S1 and S2, two regions 
southeast of S1 and S2 (labeled PS1 and PS2), 
and one region in front of (i.e. 
in the sense of propagation of the shock,
north-northwest of) S1 and S2 (labeled US1 in Table~\ref{sfits}).
The SW lobe is commonly believed to lie
behind the plane of the sky containing the nucleus
\citep{isr98, tin98}, so that any line
of sight through regions PS1 and PS2 passes through unshocked ISM,
a thin shell of shock heated gas, and the radio lobe (not visible in the X-ray
band).  We interpret regions PS1 and PS2 as dominated by unshocked ISM that lies
along the line of sight between us and the lobe.
The lobe is probably expanding spherically, and the shocked gas S1 and S2
are just breaking out of the dense gas of PS1/2 as the lobe inflates
to the north.  Any line of sight through regions PS1 or PS2
likely pass through two thin layers of the shock-heated shell,
but the path lengths through the shell
are much shorter than through the ISM, so the best fit gas
temperature is representative of the ISM.  The hot, shocked
heated shell isn't visible over the ISM through these
lines of sight.  Emission from the shocked
gas is much more prominent in S1 and S2, however,
because our line of sight through them is nearly tangent to the
shock front, maximizing its path length.

We fitted single temperature, absorbed APEC models to the spectrum of each
region.  Background was determined from a distant region.
Visual examination of archival HST/ACS data indicates that
there is absorption by cold gas in these regions, so
we allowed the value of the column density to vary freely, although
the minimum was fixed at the Galactic value.
The elemental abundance, $Z$, was held fixed at the Solar value.  
The abundance is poorly constrained if allowed to vary freely as it
can be traded off against the normalization since the emission is
line dominated.  We feel that fixing the abundance at the Solar value is
a reasonable approximation since the lobe is likely to be expanding into gas
of the merging spiral galaxy.  Since the emission is line dominated,
the proton density, $n_p$, of these features scales as $\sim Z^{-1/2}$.
The results of the fits for all five regions are summarized in Table~\ref{sfits}.
The spectral fits show a clear jump in temperature at S1 and S2,
compared to US1, PS1, or PS2, conclusively demonstrating that these
surface brightness discontinuities are due to gas that has been heated and
compressed as a result of crossing a shock front.

\begin{figure}
\plotone{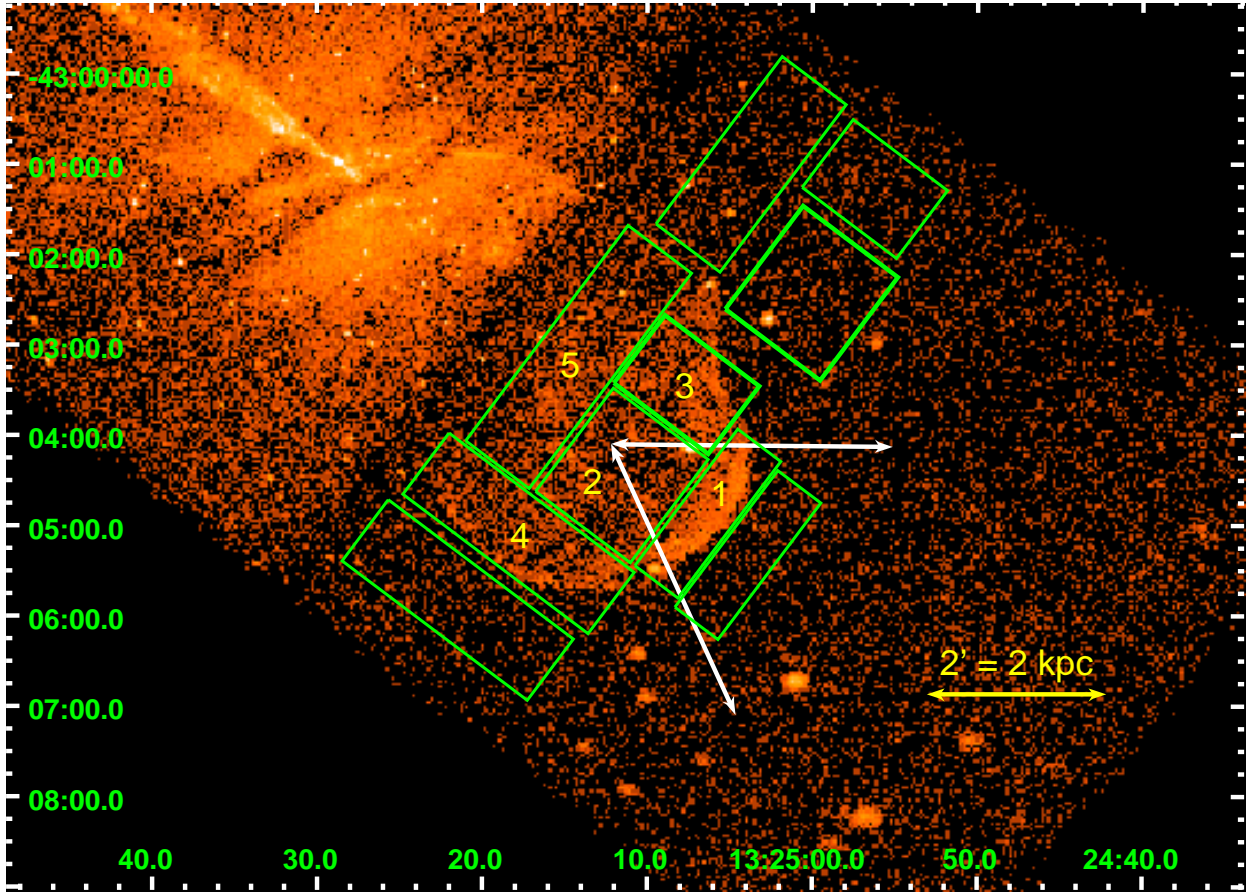}
\caption{Raw X-ray image of the southwest radio lobe of Centaurus A in
the 0.5-2.0 keV band showing regions used for spectral
fitting and background subtraction.  Region 1 was also divided
into three radial subregions (1a, 1b, and 1c) as discussed in the
text.  The best-fit values of the fitted parameters and uncertainties
are contained in Table~\ref{fittab}.  The white lines denote the
approximate position of the surface brightness wedge shown in
Figure~\ref{sbwedge}.}\label{specregs}
\end{figure}

\begin{figure}
\plotone{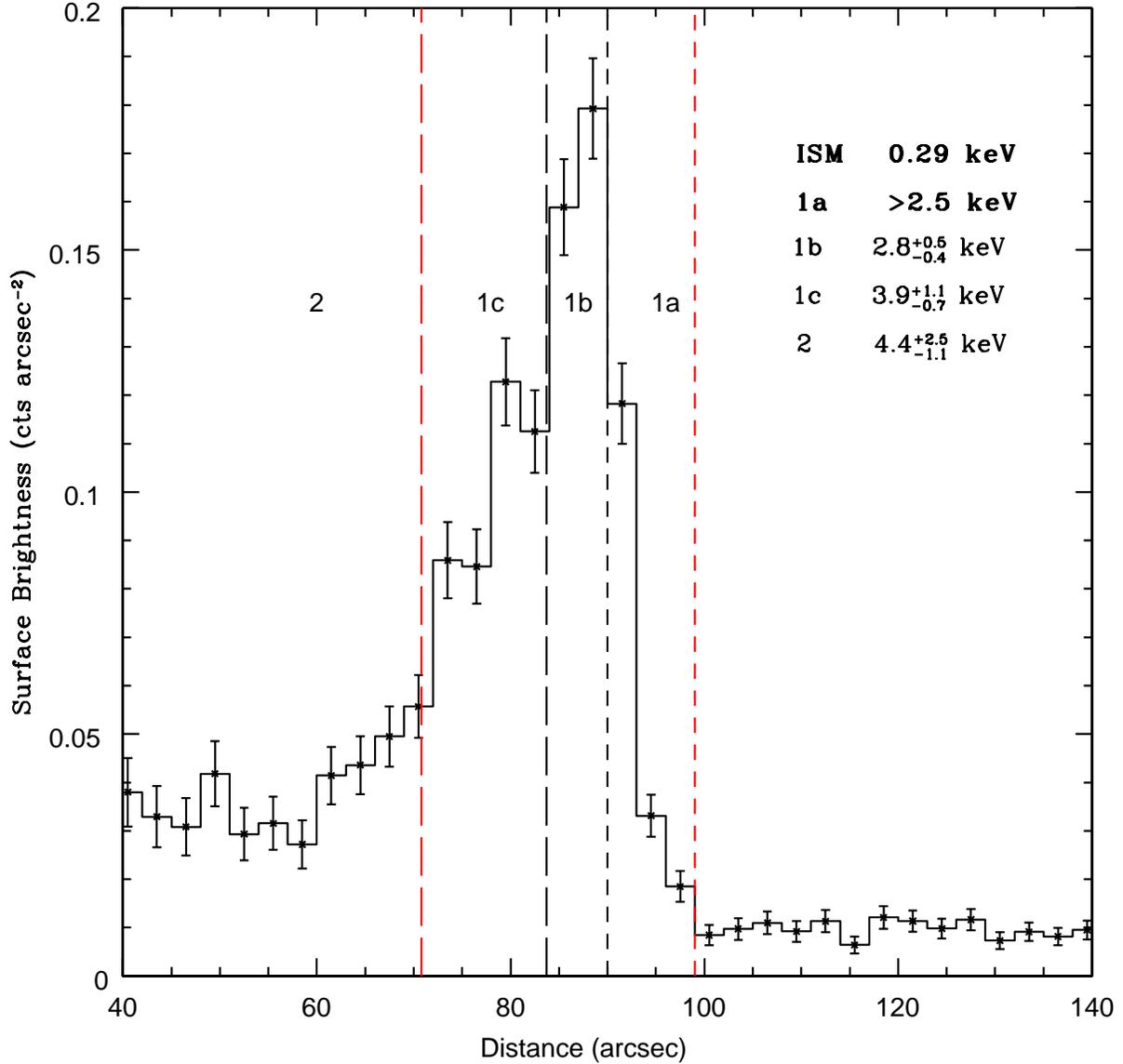}
\caption{Surface brightness profile from the center of the southwest
lobe in a 60$^\circ$ sector toward the X-ray enhancement along
the southwest boundary in the 0.5-2.0 keV band.  The region between
the two red dashed lines is the approximate thickness of the shock
(i.e. the distance between the shock and contact discontinuity defined
by the edge of the radio lobe - $\sim$28$''$).  The regions 1a, 1b, and 1c used for spectral
analysis are also identified.  Error bars on the data points are 1$\sigma$ uncertainties
due to counting statistics.  The best fit temperatures and 90\% uncertainties
for each of the regions is summarized on the right (see Table~\ref{fittab}}\label{sbwedge}
\end{figure}

For spectral analysis on larger scales, we divided the southwest radio lobe 
into the five regions shown in Figure~\ref{specregs}.  The bright
enhancement at the southwest boundary of the lobe, region 1, has been subdivided
further into three regions, referred to as 1a, 1b, and 1c, for spectral
analysis.  Figure~\ref{sbwedge} contains a plot of the surface brightness profile
of the southwest lobe in a 60$^\circ$ sector centered on the lobe.
The regions 1a, 1b, and 1c are shown.
We fit the spectrum of each region using a single temperature
APEC model with Galactic ($N_H$=8$\times$10$^{20}$ cm$^{-2}$) absorption
and fixed the elemental abundance, $Z$, at 0.5 times the Solar value.
Again, the elemental abundance is poorly constrained if allowed to vary freely.
Unlike the interior region, however, the derived proton densities are only a weak
function of the elemental abundance as the emission is continuum dominated.
We chose a lower value for the abundance here as the lobe is expanding 
into gas of the elliptical galaxy that is unlikely to have been enriched/contaminated 
by the merging spiral galaxy.
The best-fit temperatures and 90\% uncertainties for our spectral fits
are contained in Table~\ref{fittab}.
Local background was determined near the lobe.
We restricted the energy band of the fit to 0.5 and 3.0 keV in
order to minimize contamination
from the wings of the PSF of the bright nucleus (which dominates the
background above 3 keV over most of the FOV), although
our results are statistically unchanged if the fit bands are extended to 5 keV.

Along the periphery of the lobe, the single temperature fits for regions 3 through 5 are
poor with significant residuals seen between 0.6 and 1.0 keV, the Fe L shell 
complex of emission lines.  We fit these data with two temperature 
APEC models (with Galactic absorption), 
and while the fits are improved, the error bars are so large that no definitive
conclusions can be drawn.
This suggests that the emission-line temperature may be somewhat less
than the continuum electron temperature (i.e. that the electrons have not
thermalized with the ions and have not reached collisional ionization equilibrium).
We also fitted these data with a non-equilibrium ionization model
(`nei' in XSPEC 12.0) with the elemental abundances fixed at 0.5 times
Solar.  The fits were not greatly improved, and still formally unacceptable.
We conclude that the spectra of regions 3 through 5 are not well described 
by single temperature thermal models, but multi-temperature and non-equilibrium
ionization model provide little improvement.  This suggests a temperature
and ionization structure that is too complex to be resolved using the existing data.

\subsection{NE Radio Lobe}

\begin{figure}
\plotone{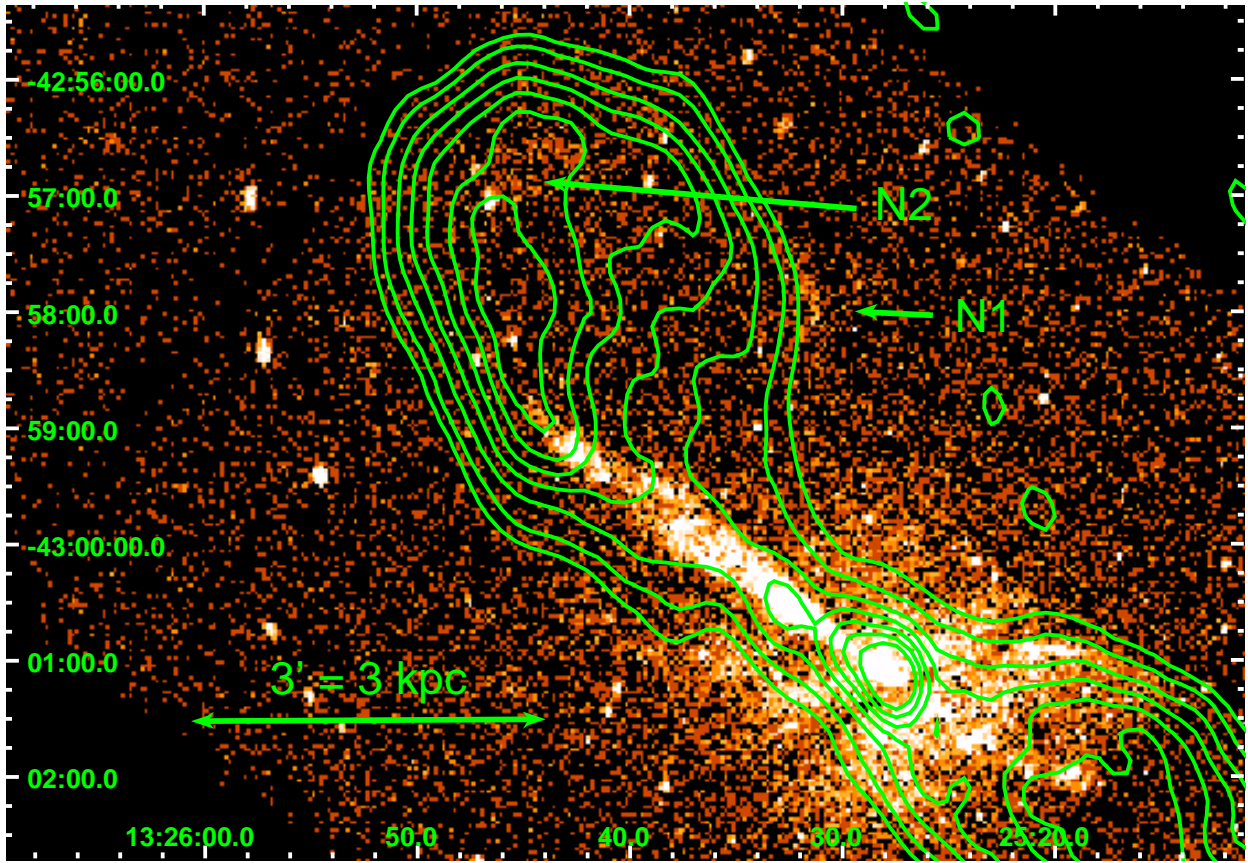}
\caption{Raw X-ray image (ACIS-S, 0.5-2.0 keV bandpass)
of Centaurus A with 13 cm radio contours (beam 30.4$''$$\times$20.3$''$
FWHM) overlaid.  Two X-ray enhancements described in the text are labeled N1
and N2.}\label{nelobe}
\end{figure}

We also detect sharp surface brightness discontinuities associated
with the NE radio lobe.  Two arcs of X-ray emission,
labeled N1 and N2 in Figure~\ref{nelobe}, are located along
the periphery (N1) and the interior in projection (N2) of the lobe.  
The morphology and location of N1 suggests a shock which would
imply that the NE lobe is expanding supersonically into the ISM, similar
to the SW lobe.  The minimum pressure of the NE lobe greatly exceeds any
plausible pressure of the ambient ISM.
This conclusion is less clear for N2 as it overlies the lobe in projection.
The spectra of both regions are poorly fit by single temperature
APEC models, although there is considerable flux above 1 keV
in both, implying gas temperatures $>$1 keV.

It is surprising that the radio morphologies and minimum pressures of the
NE and SW lobes are so similar, but their effect on the ambient ISM is so different.
Infrared synchrotron emission has been detected from the NE
lobe \citep{bro06,mjh06}, so the jet is still actively accelerating
particles to relativistic velocities in this lobe.
Several compact X-ray and radio knots in the SW lobe strongly suggest
collimated flow in this direction as well, even if there are no
structures that we can definitively term a jet \citep{mjh03}.
The one significant difference between the lobes is that
the NE lobe appears to be connected to the Northern Middle Lobe (NML) through the large-scale
jet \citep{mor99}.  How in detail the inner jet, NE lobe, large scale jet, and
NML are related is unclear, but it is almost certainly connected to why we don't
see a bright, strong shock around the NE lobe.
In particular, the energy and momentum of both the jet and the counterjet must
be comparable (otherwise the jet would push the SMBH out of the nucleus on a
short timescale).  However, the collimated flow from the AGN to the SW
is inflating a hot bubble, whereas the flow to the NE is travelling almost
losslessly (the luminosity of the X-ray jet is small compared with the mechanical
power of the jet) beyond the NE lobe.

\section{Interpretation}

\subsection{Temperature Structure}

\subsubsection{Northern periphery of SW lobe}

The sharp surface brightness discontinuity and the temperature jump
at regions S1 and S2 and in the SW demonstrate that the lobe is
expanding supersonically in the plane of the sky, and hence
is likely to be expanding supersonically in all directions.
The velocity of the shocks between S1/S2 and the undisturbed
ISM can be estimated from the ratio of the
pre-shock to post-shock temperatures.  It is not clear if
regions PS1/PS2 or US1 should be used to determine the
thermodynamic parameters of the unshocked gas.  The complex
morphology of the X-ray surface brightness, combined with spatial
variability in the unshocked gas temperature and absorption and
uncertainties in the three dimensional distribution of the gas
make determination of the density profile virtually impossible.
However, the gas temperatures of PS1, PS2, and US1 are
identical, so we can make some quantitative statements about
the energetics and dynamics of the shocks without full knowledge of
the density profile.

The ratio of post-shock to pre-shock gas temperatures, $T_R$, as a function
of Mach number is (for a purely hydrodynamic shock and $\gamma$=5/3)
\begin{equation}
T_R=T_2/T_1=(5M^2_1-1)(M^2_1+3)/(16M^2_1),
\end{equation}
where $T_2$ and $T_1$ are the post and pre-shock gas temperatures,
respectively, and $M_1$ is the Mach number of the flow in the pre-shocked
gas.  These temperature ratios are 
2.7$\pm$0.5 and 2.8$\pm$0.7 for regions S1 and S2, respectively,
at 90\% confidence assuming the pre-shock gas temperatures of PS1 and PS2, respectively.
The Mach numbers are then 2.4$\pm$0.3 and 2.5$\pm$0.5.
The uncertainties on the Mach numbers are large because the fractional
uncertainties on the pre-shock gas temperature are large.
The velocities of the shocks S1/S2 and the undisturbed ISM
are $\sim$600$\pm$75 and 680$\pm$140 km s$^{-1}$, respectively.
Assuming that the pressure of the lobe is uniform (a good assumption
as the sound speed of the lobe plasma is likely orders of magnitude larger than the
thermal gas), the ratio of the pre-shock density of S1 to that of S2 is 1.3$\pm$0.3 based
on their relative surface brightnesses.
We point out that the lower shock temperature of PS1 and PS2 relative
to the regions more distant from the nucleus (1 through 5) is
also qualitatively consistent with a picture where the nearly isobaric lobe
is expanding more slowly into the denser regions of gas near the nucleus.

\subsubsection{SW periphery of SW lobe}

As discussed in \citet{kra03}, we model the emission as a shell of
uniform density rotated to our line of sight.  In Figure~\ref{sbwedge},
we have labeled the approximate positions of the contact discontinuity between the shocked
gas and radio lobe (the red dashed line on the left), the shock-heated shell (the
region between the two dashed red lines), and the transition region (the actual thickness
of the shock, see below for detailed discussion - labeled 1a).
We estimate the thickness of the shell and the transition 
region to be $\sim$28$''$ (476 pc) and $\sim$9$''$ (153 pc), respectively.
The distance from the shock to the contact discontinuity is therefore $\sim$37$''$ (630 pc).
The width of the transition region is estimated as the distance over which
the surface brightness of the shell goes from the background level to
its peak value.  This is an upper limit on the actual thickness of the transition region
as we have neglected projection effects. 
The ratio of the gas temperature in region 1b to region 1c
is 0.72$\pm$0.20 (90\% confidence).
Thus the temperature of the gas closest to the shock is cooler (at marginal
significance) than the gas behind the shock closer to the lobe.
We have neglected the effects of projection, but projection would tend
to wash out any temperature differences, so our estimate of the temperature
ratio is really an upper limit.
The temperature of the material in the transition region (region 1a) 
is poorly constrained, but is $>$2.5 keV at 90\% confidence.
Thus the transition region is not significantly cooler than regions 1b or 1c.
A detailed map of the temperature structure of this shock-heated
shell would permit us to make a strong statement about the limits of
the applicability of a purely hydrodynamical model to the lobe/ISM
interaction as we argue below.  

\subsection{Transport Processes and Electron-Ion Equilibration in the Shock around
the SW Radio Lobe}

\subsubsection{Theoretical Considerations}

\begin{figure}
\plotone{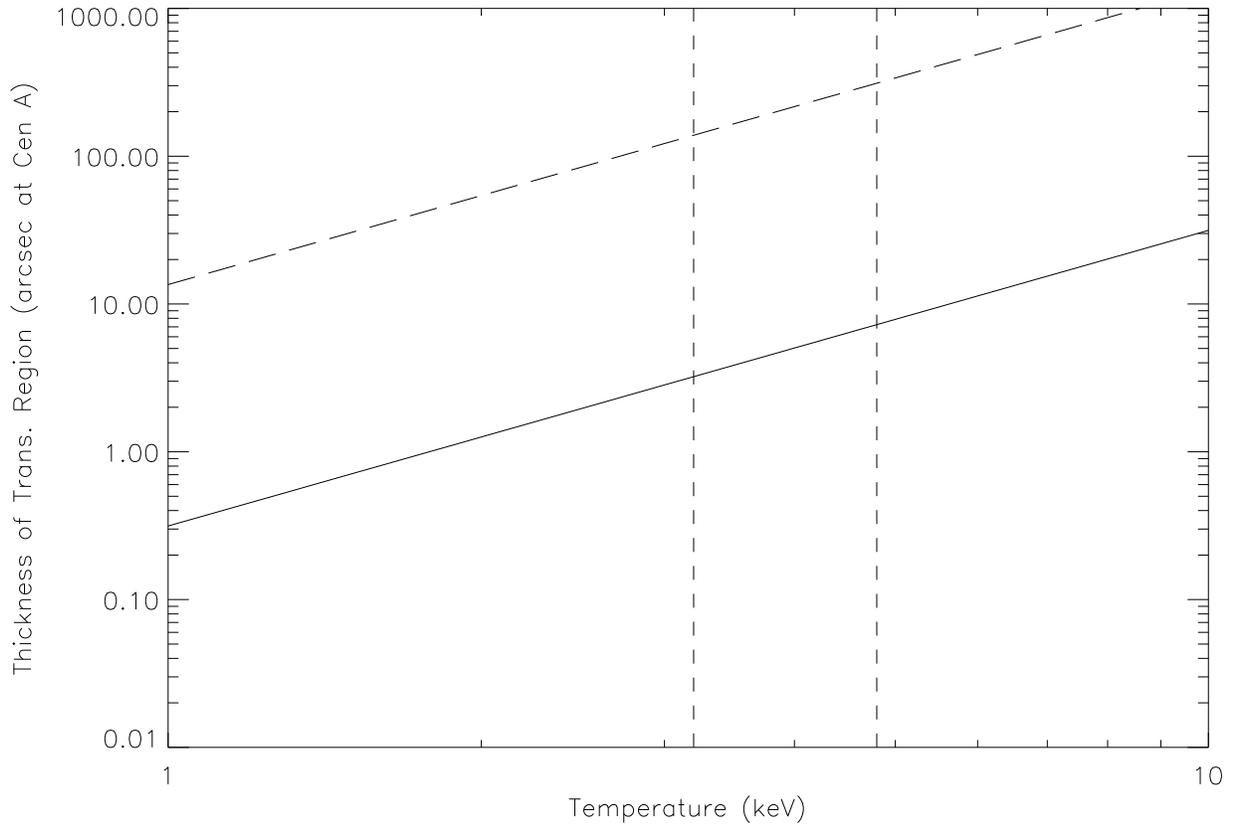}
\caption{Plot of thickness of ion shock for ion-ion collisions
(solid curve) and ion-electron collision equilibration length (dashed curve)
for shock around the southwest radio lobe of Centaurus A assuming
the energy transfer between particles is governed by Coulomb
collisions at the Spitzer rates.  The vertical
lines denote the upper and lower limits (90\% confidence) of the
temperature of the shell.}\label{treg}
\end{figure}

It is almost always assumed that the physics governing radio lobe/ICM interactions
is purely hydrodynamic.
That is, the electron-ion plasma can be considered as a single, classical fluid.
This may not be a good approximation for the high Mach number shock
around the SW radio lobe of Cen A because of its proximity, temperature, and density.
The mean free path, $\lambda_{ii}$, for collisional energy exchange
between the ions (protons) is
\begin{equation}
\lambda_{ii} = 230 pc \times (T_i/10^7 K)^{2} (n_p /10^{-3} cm^{-3})^{-1},
\end{equation}
where $T_i$ and $n_p$ are the ion temperature and density, respectively \citep{spi62}.  
For simplicity we assume that the plasma is pure hydrogen.
The great bulk of the gas kinetic energy is carried into the
shock by the ions.  In a strong, collision-dominated shock the
kinetic energy is thermalized among the ions over a distance comparable to $\lambda_{ii}$.
Collisional energy exchange between the ions and
electrons is a factor of $\sim \sqrt{m_p/m_e} \simeq 43$ slower, so that
the region over which the electron temperature differs significantly
from the ion temperature is roughly 40 times larger than the thickness of
the ion shock.  At the observed temperature of $\sim$3.5-4.0 keV, the ion shock
around the SW radio lobe of Cen A should be spatially resolvable.
A plot of the predicted thickness of the transition region as a function of post-shock
temperature is shown in Figure~\ref{treg} for the measured gas density
($n_p=2.2\times 10^{-2}$ cm$^{-3}$) of the shell.  The temperature of the gas in the
southwest region of the shell (Region 1 of Table~\ref{fittab})
is 3.9$\pm$0.7 keV (for $Z$=0.5, 90\%
confidence).  The region of allowed parameter space
for ion-ion (solid) and ion-electron collisions (dashed) is
denoted by the vertical dashed lines in Figure~\ref{treg}.
Thus the thickness of the ion shock around the
lobe would be several arcseconds at the distance of Cen A.
The distance scale for ion-electron equilibration is also shown in Figure~\ref{treg}.
Around the SW lobe in Cen A, this would be more than 1 kpc, which is larger than
the thickness of the shell.

Observations of young Galactic and Magellanic supernova remnants (SNRs)
demonstrate that the ion shocks are collisionless \citep{rak05},
therefore the ion shock in Cen A is
likely to be orders of magnitude smaller than estimated from ion-ion collisions.
Plasma effects and magnetic fields, even if not dynamically
important, can reduce the mean free path for energy and momentum transfer between {\em ions}
to a value many orders of magnitude smaller than the Spitzer estimate.
The ion shock of Galactic supernova remnants ($\sim$1000 times closer than
Cen A) with gas temperatures similar to the shock-heated shell in Cen A
have never been spatially resolved.
It would therefore be surprising if we could observe this region in Cen A.

However, the efficiency with which the protons transfer energy to the electrons in
SNR shocks (and in low density, high Mach number plasma shocks in general) is largely unknown.
There may be some collisionless heating of the electrons in the ion shock, but
it is believed that this heating will not be efficient and that the electron
temperature will be significantly below the ion temperature at the boundary
of the ion shock (i.e. where the protons reach their final, post-shock
temperature) \citep{bag87,car88,sch88}.
The plasma (i.e. wave-particle interaction)
and MHD processes that reduce $\lambda_{ii}$ in the ion shock of
SNR shocks do not appear to greatly reduce $\lambda_{ie}$.  In fact,
large differences between the electron and ion temperatures have been
measured in several young SNRs including SN 1006 \citep{vink03}, Tycho, the
Cygnus Loop \citep{ray03}, and the LMC remnant Dem L71 \citep{rak03}.
Comparison of X-ray measurements of electron temperatures, $T_e$, with
H$_\alpha$/H$_\beta$ line ratio estimates of the ion
temperature, $T_i$, indicates that there is a strong correlation between
the shock velocity and the ratio of $T_e$ to $T_i$ \citep{rak05}.
Stronger shocks in young SNRs tend to have lower ratios of $T_e$/$T_i$.
For Cen A, the ratio of the temperature of the gas in the shell ($\sim$3.5 keV) to the
ISM (0.3 keV) is $\sim$12.  This temperature ratio implies a shock
velocity (for a purely hydrodynamical shock) of $\sim$1500 km s$^{-1}$
($M\sim$6.2).  For SNR with a similar shock velocity such as Tycho, $T_e$/$T_i\sim$0.2.
Therefore, if the transport processes relevant to the expansion of the
SW radio lobe of Cen A are similar to those in young SNRs,
it is likely that the proton temperature is considerably
higher than the electron temperature, and the electron temperature
(i.e. the temperature that we measure with the X-ray spectrum)
considerably underestimates the strength of the shock.

Therefore, the electrons and ions are unlikely to have reached thermal equilibrium 
and there should be an observable radial temperature gradient in the shell.  
Additionally, since we measure the electron (and ionization)
temperature with the X-ray spectrum, it is likely that we have underestimated the 
ion temperature and shock velocity, so the shock may be even stronger than we estimate
based on the electron temperature.  
A time-dependent consideration of Coulomb collisions in plasmas
suggests that the ratio, $q$, of the electron temperature, $T_e$, to
the final (equilibration) temperature $T_f$, after time $t$ is given by
\begin{equation}
dq/dt=Kq^{-3/2}(1-q),
\end{equation}
where K=2.75$\times$10$^{-4}$$n$$(T_f/10^7K)^{-3/2}$ yrs$^{-1}$ and
$n$ is the total (i.e. $n_e+n_i$) particle density \citep{spi62}.
Thus the electron temperature would rise to roughly half
the ion temperature in a few times the Spitzer ion-ion collision length (tens
of arcseconds in our case), then more slowly approach equilibrium 
over a distance 43$\times \lambda _{ii}$ 

\subsubsection{Simulations}

To evaluate this phenomenon quantitatively, we created a one dimensional
spherical shock model in a two fluid (electron/proton) plasma
driven by energy injected from the center.
Several simplifying assumptions have been made.  First,
we assume energy is transferred between the particles only
by Coulomb collisions, and that the rate of energy transfer
is given by the Spitzer value.
Second, we assume that there is no separation between the
electrons and ions (i.e. $n_e$=$n_i$).  This latter approximation
is extremely good as the maximum length scale of separation is on the
order of the Debye length, which is hundreds
of meters for the parameters of the Cen A shock.
Third, we introduce an artificial proton viscosity 
(the Richtmyer-Morton artificial viscosity) to
ensure that we capture the features of the shock at
the resolution of the simulation.
Finally, we neglect the effects of thermal conduction.
More detailed studies of two-fluid shocks demonstrate that
thermal conduction from the downstream electrons
can heat the pre-shock electrons, thus creating a shock
precursor \citep{cas91}.  The presence of such a precursor
has not been seen in Galactic SNRs, and is unobservable in our data.
We emphasize that we are interested in studying the thermal
relaxation between the ions and electrons, not thermal conduction.

Under these conditions, motion of the two fluids is described by a
single continuity equation
\begin{equation}
{d\rho\over dt} + \rho \nabla \cdot {\bf v} = 0,
\end{equation}
where $\rho$ is the total density and ${\bf v}$ is the common velocity
of the two fluids.  The lagrangian time derivative has its usual
meaning, $d/dt = \partial/ \partial t + {\bf v} \cdot \nabla$.  The
single momentum equation is
\begin{equation}
\rho {d{\bf v} \over dt} = - \nabla p + \nabla \cdot {\bf T} + \rho{\bf g},
\end{equation}
where $p$ is the total gas pressure, ${\bf T}$ is the viscous stress
tensor, and ${\bf g}$ is the acceleration due to gravity.  Only the
artificial viscosity contributes to the viscous stresses in the
simulation.  The energy equation for the protons is
\begin{equation}
\rho_{\rm p} {d\epsilon_{\rm p} \over dt} = {p_{\rm p} \over \rho_{\rm p}} {d\rho_{\rm p} \over dt} + \Pi_{\rm visc} + \xi_{\rm pe}
\end{equation}
and that of the electrons is
\begin{equation}
\rho_{\rm e} { d\epsilon_{\rm e} \over dt} = {p_{\rm e} \over \rho_{\rm e}} {d\rho_{\rm e} \over dt} + \xi_{\rm ep}.
\end{equation}
Here $\rho_\alpha$ is the density of a fluid component ($\alpha =
\rm p$ or e), $p_\alpha$ is the corresponding component of the
pressure and $\epsilon_\alpha$ is the specific thermal energy
($\epsilon_\alpha = \gamma p_\alpha / [(\gamma - 1) \rho_\alpha]$).
By our assumptions, the viscous heating rate, $\Pi_{\rm visc}$, only
affects the protons directly.  The rate of energy transfer between the
fluids is
\begin{equation}
\xi_{\alpha\alpha'} = - n k_{\rm B} (T_\alpha - T_{\alpha'}) /
\tau,
\end{equation}
where the equilibration time $\tau$ is
\begin{equation}
\tau = \frac{3 m_em_ic^3}{8\sqrt{(2\pi)}e^4n_e{\rm ln}\Lambda} \sim {\rm 9300\ yrs} \times \frac{T^{3/2}_e(keV)}{n_e(cm^{-3})}.
\end{equation}
The factor ${\rm ln}(\Lambda)$ in the denominator of equation 9 is the
Coulomb logarithm and has a weak (logarithmic) dependence on the temperature.
The relative thermal speed, $c$, of the particles 
($(\frac{k_BT_p}{m_p}+\frac{k_BT_e}{m_e})^{1/2}$) is dominated by the
electrons at the electron and ion temperatures of interest in this paper.
Finally, we assume an ideal gas equation of state for both the electrons
and protons with $\gamma$=$\frac{5}{3}$.

\begin{figure}
\plotone{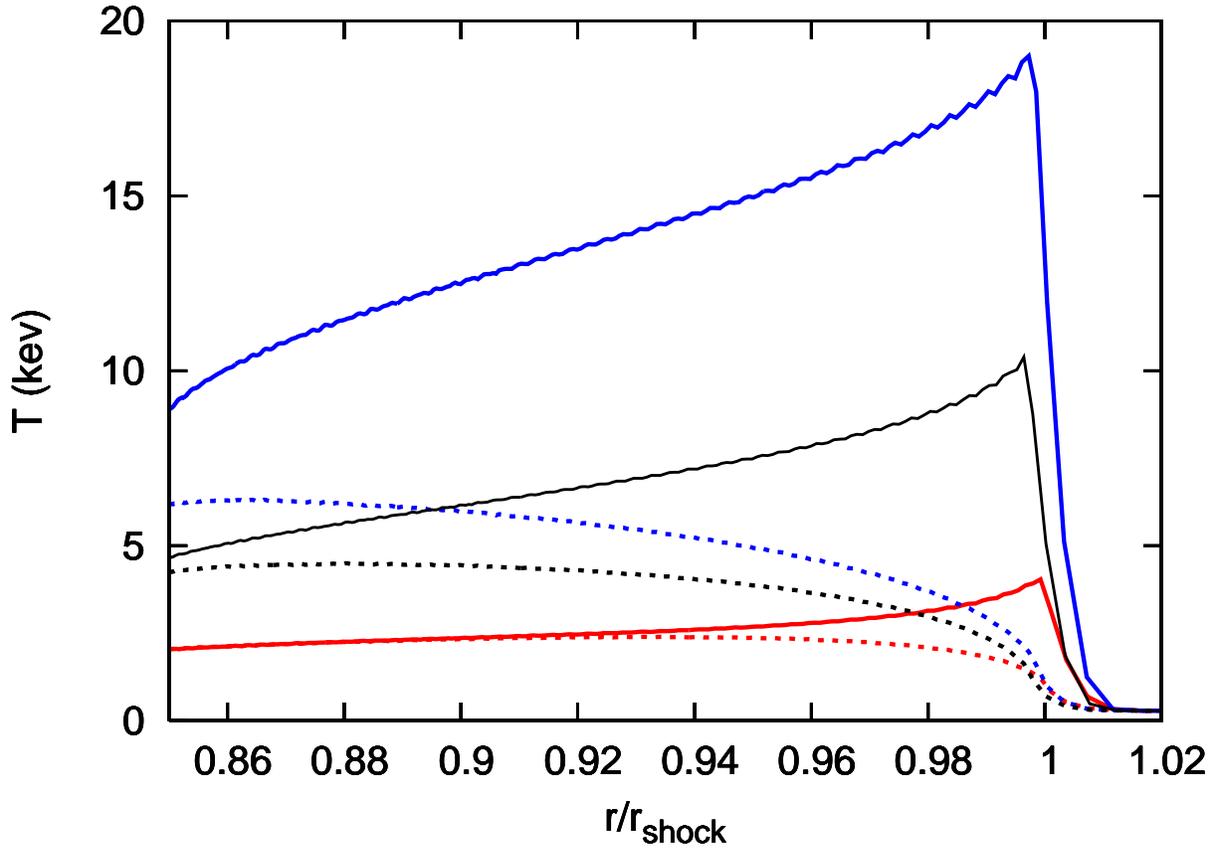}
\caption{The proton (solid lines) and electron (dashed lines) temperatures
as a function of distance behind the shock of the two fluid
shock model for three Mach numbers.  The blue, black, and red curves
correspond to shock Mach numbers of 10.8, 8.3, and 5.2, respectively.}\label{shocktemp}
\end{figure}

We simulate the shock as a continuous release of energy at the center of
an isothermal atmosphere with a power law density distribution.
The parameters of the ambient gas were matched to
measurements of Cen A ($\beta$=0.40, $k_BT$=0.29 keV, see \citet{kra03}
for details).  In our earlier paper, we found that the density jump
at the shock was roughly a factor of 10, much larger than the factor
of 4 required by the Rankine-Hugoniot relations for a strong shock
in a fluid with $\gamma$=5/3.  This large density jump cannot be accounted
for in our simulations, and would require the inclusion of additional
physics (e.g. the creation of cosmic rays at the shock).
Conservatively, we set the ambient ISM to a higher density so that the post-shock
value matches the measured density of the shell.
If the density of the ISM and shell is, in fact, lower, our conclusions
are strengthened as the timescales for equilibrium are even longer than
in the simulations presented in this paper.
Three values of the initial input energy were chosen.  In the
first simulation, the energy was chosen such that the Mach number
($M$=8.3) of the flow at the distance of the SW boundary of the lobe from
the nucleus ($\sim$6.5 kpc) matches the value quoted by \citet{kra03}.
Two other values were chosen, $M$=5.2 and 10.8, to bracket this choice of Mach number.
The electron temperature as a function of distance behind the shock
for each of the three cases is shown in Figure~\ref{shocktemp}.

In all these simulations, the proton temperature rises rapidly to roughly twice its
final value, since all of the kinetic energy of the shock is initially
transferred to the protons.  The electron temperature then begins
to rise rapidly, but levels off at about 3.5 keV (in the
Mach 8.3 case), roughly 40\% of the proton temperature.  As the electron
and proton temperatures approach each other, the rate of energy transfer
decreases, ultimately being balanced by adiabatic losses as the plasma expands after
passage of the shock.  Thus the electron temperature reaches a plateau with very
little gradient from the shock to the contact discontinuity.  There is,
however, a significant gradient in the proton temperature between
the shock and the contact discontinuity.
For the spherical model used here, all gas initially
outside the cavity remains in place as the cavity expands.  In a more
realistic model, the shocked gas may flow around the radio lobe as
the lobe pushes outward if the expansion of the lobe is largely
radial from the nucleus.  As a result, the shocked gas closest to the
cavity would have flowed away and the remaining shell of shocked gas
would be thinner than for the model.
This is probably not significant for the Cen A lobe as it appears to
be overpressured, and hence expanding supersonically, around the
entire periphery.

\subsubsection{Implications}

Consideration of the transport processes thus
has several important implications in the case of Cen A.
First, it suggests that the proton temperature is considerably
higher than the electron temperature at the strongest part
of the shock.  Thus our earlier estimates of the total power
of the lobe expansion may be low by a factor of up to a few.
Second, the observed electron temperature is not a
sensitive diagnostic of the shock velocity or the
energy in the shock.
In the three cases shown in Figure~\ref{shocktemp}, the post shock
electron temperature varies by only a factor of $\sim$2, while
the initial post-shock proton temperature varies by more than a
factor of 4.  

Third, our simulations predict that there will be little 
temperature structure in the shell
between the shock and the contact discontinuity
except for a small region of lower temperature just behind the shock.
Purely hydrodynamic simulations of a high Mach number shock around
a solid sphere in a uniform density atmosphere
show a $\sim$10\% increase in the gas temperature
from the shock to the contact discontinuity.  This can be easily
demonstrated from Bernoulli's equation, the non-zero velocity
of the gas just behind the shock must be converted to thermal
energy of the gas at the contact discontinuity.  The Sedov solution
for a point release of energy in an atmosphere with a power law
density gradient predicts an even larger gradient.
\citet{kai99} describe this process for a range of model atmospheres.
Data of sufficient quality should be able to clearly distinguish between
these alternatives.

\begin{figure}
\plotone{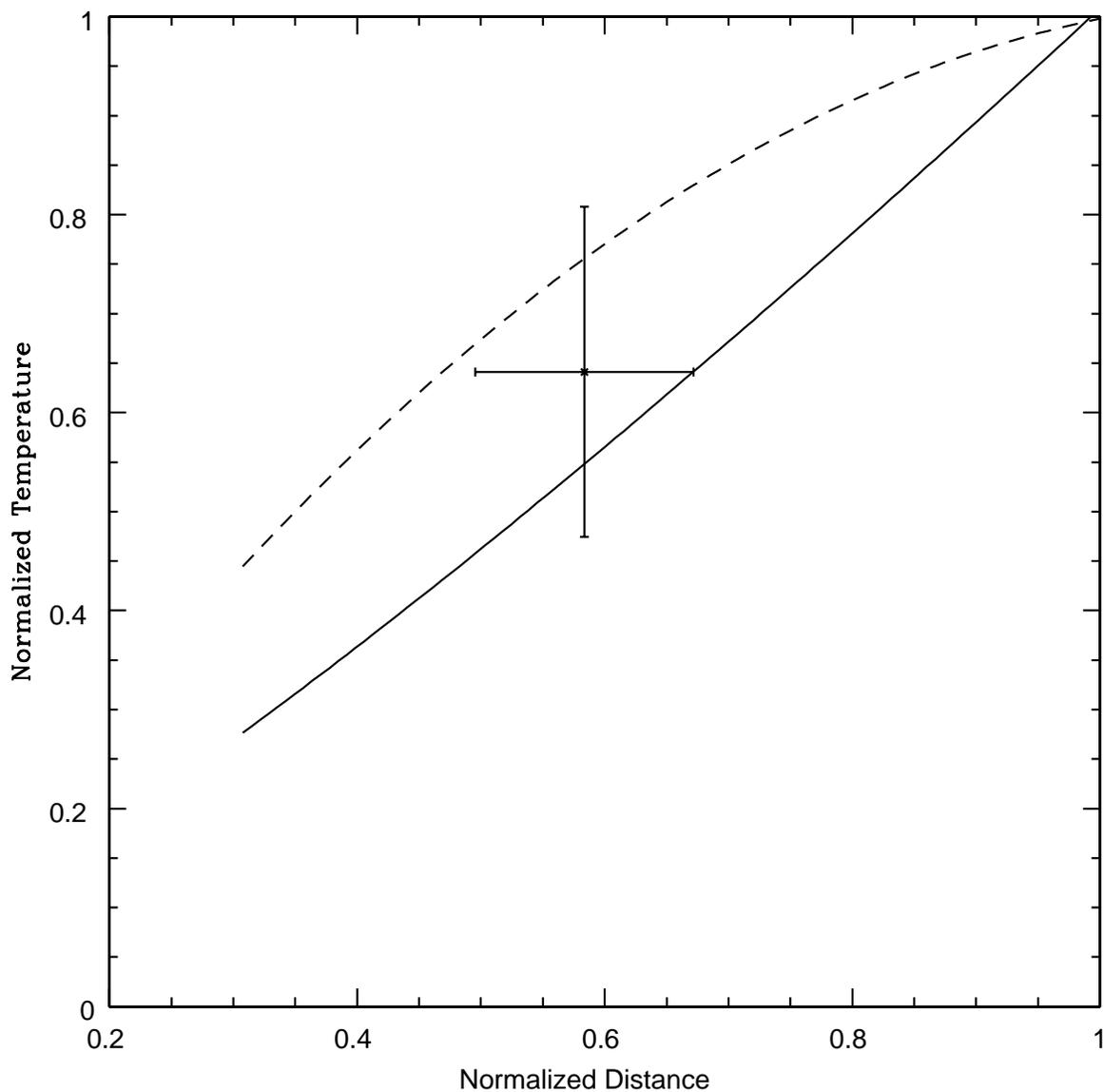}
\caption{Plot of the electron temperature of the shock (normalized to
the observed value at the SW boundary of the lobe) as a function
of distance from the nucleus (normalized to the distance from the
nucleus to the leading edge of the shock) for $M$=8.3 at the
leading edge. The solid curve is the prediction based on a purely
hydrodynamic model of the shock, the dashed curve from our
two-fluid shock model.  Thus, this plot shows the predicted variation in
observed electron temperature of the shock around the periphery of the radio lobe.
The point with error bars is the ratio
of the observed temperatures in region 5 to region 1.  The error bars
on the y-axis are the 90\% confidence uncertainties.}\label{tequil}
\end{figure}

Fourth, there will be less temperature structure around the
periphery of the lobe than one would naively expect based on the
Rankine-Hugoniot shock conditions.  In particular, the shock
will be weaker around the sides (i.e. closer to the
nucleus) of the lobe since the nearly isobaric
lobe is expanding into denser material.  Since the shock will
be weaker, and the ambient density higher,  the electron and proton temperatures 
will equilibrate more rapidly than at the SW periphery of the lobe.
The observed electron temperature of the shell nearest the nucleus
will be close to the final temperature and a much better, though
imperfect, diagnostic of the strength of the shock.
The post-shock electron temperature (normalized to the value at
the SW edge) as a function of normalized distance between the nucleus
and the SW edge is shown in Figure~\ref{tequil}.  The solid curve
is the normalized temperature if the shock is purely hydrodynamic,
the dashed curve is for our two-fluid shock model.  There is a clear difference
in the temperature profiles.  The point with the error bars is the measured
ratio of the temperature in region 5 to that in region 1.  The existing
data are not adequate to distinguish between the two models at 90\%
confidence.

This also has important implications for similar features
in other radio galaxies and clusters of galaxies.  
Differences in the electron and proton temperatures
will make it very difficult to detect strong shocks in young, powerful
radio sources.  In the earliest stage of the development of a radio
galaxy (i.e. when the jet is momentum dominated), the shock temperature
could be tens or even hundreds of keV and the equilibration time
would be tens or hundreds of millions of years.  This is orders of
magnitude longer than the lifetime of the source in this stage.  
The proton temperature could be quite high with little or no change
in the electron temperature.
Finally, the importance of transport processes
could have important implications for the long term evolution of
powerful radio galaxies in poor environments.  In these cases,
the advance speed of the jet head can remain supersonic for tens or
hundreds of kpc.  The timescale for electron/proton equilibration
could be hundreds of millions of years (or more).  
In the absence of other processes to transfer energy from the protons to
electrons, the atmospheres of poor systems could remain far from equilibrium
for a considerable period of time.

\section{Conclusions}

The hot thermal shell of shock-heated gas surrounding the southwest radio lobe
of Centaurus A is the best example of a spatially resolvable high Mach number
shock in an extragalactic system.  It is therefore a unique
laboratory in which to study the hydrodynamics and plasma physics
of the radio lobe/ISM interaction.
We find that the surface brightness discontinuity extends
around $\sim$75\% of the boundary of the SW radio lobe.  
The shock likely extends around the entire lobe, but the current observations
do not yet have the sensitivity to detect it.
We also report the discovery of two filaments of X-ray emission associated
with the NE lobe, although the data quality is not sufficent to conclusively
determine if they are shock-heated gas.

We demonstrate that if the energy transfer between electrons
and protons behind the shock of the SW lobe is purely collisional,
their temperatures will not have equilibrated.  One dimensional,
two-fluid field-free simulations show
there will be little temperature structure in the gas between
the shock and the contact discontinuity as adiabatic losses will roughly
balance Coulomb heating of the electrons.
These simulations also predict significant differences in the temperature structure
as a function of distance from the nucleus (i.e. around the
periphery of the lobe) compared with a purely hydrodynamic model.
That is, the shock strength will vary quite strongly around the lobe
because of the density gradient in the gas.  Our two-fluid simulations
suggest that the electron temperature in the shocked gas around the lobe will be more
uniform than predicted in single fluid hydrodynamic model.

A deep ($>$500 ks) {\em Chandra} observation of Cen A is required to further
elucidate the underlying shock physics.
In particular, a deeper observation would permit an accurate measurement of the
shock temperature and pressure around the periphery of the lobe,
thus constraining both the expansion velocity, external gas pressure, and
external density.
A deeper observation of the X-ray bright enhancement at the southwest
boundary of the lobe would allow a detailed estimate of the
temperature structure in the shock.  This could then be
compared with two and three dimensional two-fluid
simulations of the shock to better estimate
the energy in the shock and the degree of coupling between the
electrons and ions.

\acknowledgements

We thank John Raymond, Cara Rakowski, and Joss Bland-Hawthorn for helpful discussions.  This work was
supported by NASA grant NAS8-01130 (the HRC GTO grant).
We also thank the anonymous referee for comments that improved this paper.

\clearpage

\clearpage

\clearpage

\clearpage 

\clearpage

\clearpage

\clearpage

\clearpage

\clearpage

\clearpage

\begin{table}
\begin{center}
\begin{tabular}{|l|c|c|c|}\hline\hline
Region & Temperature (keV) & $N_H$ (10$^{21}$ cm$^{-2}$) & $\chi^2_\nu$ \\ \hline
  S1   &   0.62$\pm$0.04   & 4.5$\pm$0.6       &  1.5  \\ \hline
 PS1   &   0.23$\pm$0.04   & 7.0$\pm$1.2       &  1.8  \\ \hline
  S2   &   0.78$\pm$0.06   & $<$2.0            &  0.84 \\ \hline
 PS2   &   0.28$\pm$0.08   & 6.9$\pm$0.2       &  1.3  \\ \hline
 US1   &   0.24$\pm$0.06   & 3.9$\pm$0.2       &  1.3  \\ \hline
\end{tabular}
\caption{Summary of best fit temperatures for regions around
S1 and S2 southwest lobe shown in Figure~\ref{xrawz}.
The value of $N_H$ includes the contribution from
Galactic material (8$\times$10$^{20}$ cm$^{-2}$).
Uncertainties are 90\% for one parameter of interest. See text
for full description of regions.}\label{sfits}
\end{center}
\end{table}

\clearpage

\begin{table}
\begin{center}
\begin{tabular}{|c|c|c|}\hline\hline
Region & Temperature (keV) & $\chi^2_\nu$ \\ \hline
  1    & 3.9$_{-0.7}^{+0.9}$ &  0.77 \\ \hline
  1a   & $>2.5$              &  0.37 \\ \hline
  1b   & 2.8$_{-0.4}^{+0.5}$ &  0.64 \\ \hline
  1c   & 3.9$_{-0.7}^{+1.1}$ &  0.81 \\ \hline
  2    & 4.4$_{-1.1}^{+2.5}$ &  1.27 \\ \hline
  3    & 3.8$_{-0.8}^{+1.4}$ &  2.39 \\ \hline
  4    & 3.1$_{-0.5}^{+0.9}$ &  1.70 \\ \hline
  5    & 2.5$_{-0.4}^{+0.6}$ &  3.02 \\ \hline
\end{tabular}
\caption{Summary of best fit temperatures for regions of southwest
lobe shown in Figure~\ref{specregs}.
Uncertainties are 90\% for one parameter of interest. See text
for full description of regions.}\label{fittab}
\end{center}
\end{table}

\end{document}